\documentclass[aps,prd,showpacs,nofootinbib,floats,floatfix,preprintnumbers,groupedaddress,twocolumn]{revtex4}

\usepackage{bm}
\usepackage{latexsym}
\usepackage{dcolumn}
\usepackage{amsmath,amsfonts,amssymb}
\usepackage{graphicx,epsfig}
\usepackage{amsmath}
\usepackage{fancyhdr}
\usepackage{hyperref}
\usepackage{graphicx,epstopdf}
\usepackage{physics}

\hypersetup{
	colorlinks   = true, 
	urlcolor     = blue, 
	linkcolor    = blue, 
	citecolor   = red 
}

\begin{document}
\title{Shock wave quantum memory in shocked detector}
\author{Bibhas Ranjan Majhi\footnote {\color{blue} bibhas.majhi@iitg.ac.in}}

\affiliation{Department of Physics, Indian Institute of Technology Guwahati, Guwahati 781039, Assam, India}

\date{\today}

\begin{abstract}
Usual uniformly accelerated frame, in Dray-'t Hooft spacetime, does not see the any quantum imprint on Unruh effect due to localised shock wave in Minkowski spacetime. Here we argue that such non-appearance of quantum memory is specific to those particular observers which do not incorporate the presence of wave in their trajectory. In fact a detector, associated with a frame which is affected by the shock, can not be trivial in terms of its response. We explicitly show that the later type of frame detects particle in the shock wave Minkowski vacuum which is the effect of shock. Therefore this quantum memory is very special to specific class of observers as far as Dray-'t Hooft spacetime is concerned. We analyse for a null like trajectory along which the detector is moving.
\end{abstract}


\maketitle

\section{Introduction}
The recent observations \cite{Hawking:2016msc,Hawking:2016sgy,Strominger:2017aeh} in the context of Bondi-Burg-Metzner-Sachs (BMS) symmetry \cite{Bondi:1962px,Sachs:1962wk,Sachs:1962zza,Newman:1966ub} shows that the asymptotic symmetry of a spacetime leads to supertranslational and superrotational charges, known as `soft hair'. It implies that the classical soft hairs are observables and keep their imprint by modifying the background metric. A general believe is that the corresponding emerged ``{\it quantum memory}'' may play a pivotal role in solving the {\it information paradox} issue in the context of black hole evaporation. In this regard several investigations have been done starting from asymptotic symmetries of spacetime \cite{Donnay:2015abr,Eling:2016qvx,Akhmedov:2017ftb,Maitra:2018saa} to its consequences at the classical as well as quantum level \cite{Zhang:2017geq,Gomez:2017ioy,Kolekar:2017tge,Donnay:2018ckb,Chu:2018tzu,Javadinazhed:2018mle,Maitra:2019eix}.

Till now it has been observed that the Hawking temperature is not being modified for the classically supertranslated Schwarzschild black hole \cite{Hawking:2016sgy,Javadinazhed:2018mle,Lin:2020gva}. Whereas the same for Vaidya black is being  changed \cite{Chu:2018tzu,Chiang:2020lem}. Therefore the quantum imprint of classical memory seems to be very specific to certain situation, but not a general phenomenon. In this regard, another very popular case can be mentioned here due to Dray and 't Hooft \cite{Dray:1984ha}. They showed that the Minkowski spacetime is modified in the presence of a localised shock wave in the spacetime and a massless particle's trajectory get shifted in the direction of the propagation of the wave. This is the signature of the classical memory effect of the spacetime. Then the next question is how quantum observables can be affected. A rigorous analysis revealed that the the particle number seen in the shock wave Minkowski vacuum (SWMV) as well as the Unruh temperature \cite{Unruh:1976db} seen by an accelerated observer is unaffected \cite{Compere:2019rof}. The same has also been confirmed in \cite{Majhi:2020pps} by evaluating the Wightman function for massless scalar field. It was observed that the Wightman function does not modify. In fact it has been shown in the same work \cite{Majhi:2020pps} that the vacuum is equivalent on both sides of the localization of the shock wave.  On the other hand the non-vacuum states of scalar field does capture the quantum imprint of this shock.

In the light of these existing results, here we want to investigate if the quantum memory is purely observer dependent phenomenon or depends on the choice of the field states. Particularly whether this effect is observer dependent even for vacuum states. We already mentioned in the last paragraph that an accelerated observer does not see this signature in the quantum regime, provided it is investigating SWMV. Dose it due to the fact that the observer is classically blind to presence of shock in the spacetime? To answer this question, in the letter a particular {\it null like} observer is chosen whose trajectory  captures the properties of shock. The path of this frame is chosen in such a way that it incorporates a shift due to shock wave. Here we investigate how a two level atomic detector, attached with this frame, responds when it interacts with the Minkowski scalar modes. The movement of the detector is confined within a region where the scalar mode also captures the signature of shock in order to investigate the solo effect of the wave. We find that the detector undergoes a transition from ground state to excited state. Moreover, the transition probability depends on the strength of the shock and thereby providing a direct evidence of the presence of shock in spacetime. Therefore the observer captures the quantum memory of classical shock wave, whereas the usual accelerated observer does not.  So we conclude that in this case {\it the quantum memory effect is an observer dependent phenomenon}.  

\section{Massless scalar modes on shock wave metric}
The back reaction of a shock wave, propagating along $X$-axis on the Minkowski spacetime which is localised at $U=U_0$, affects the background metric. If the shock wave stress-tensor is represented by $T_{UU}=\delta(U-U_0)T({\bf x}_{\perp})$, then for this case such modified metric is given by \cite{Dray:1984ha}
\begin{equation}
ds^2 = -dUdV+f({\bf{x}}_{\perp})\delta(U-U_0)dU^2 + d{\bf{x}}_{\perp}^2~,
\label{2.01}
\end{equation}
where $U=T-X$, $V=T+X$ are ingoing and outgoing null coordinates, respectively. ${\bf x}_{\perp}$ are the transverse coordinates, which will be here denoted as $Y$ and $Z$. Here $T$ satisfies $\nabla_{\perp}^2 f({\bf x}_{\perp}) = -16\pi G T({\bf x}_{\perp})$ and is function of only transverse coordinates. $\delta(U-U_0)$ is the Dirac-delta distribution.
In a shifted coordinates
\begin{eqnarray}
&&\hat{U} = \hat{T}-\hat{X} = U~; 
\nonumber
\\
&&\hat{V} = \hat{T}+\hat{X} = V-\Theta(U-U_0)f({\bf x}_{\perp})~;
\label{BRM2}
\end{eqnarray}
the metric (\ref{2.01}) takes the following form
\begin{equation}
ds^2 = -d{U}d\hat{V}-\Theta(U-U_0)\partial_Af({\bf{x}}_{\perp}) dUdx^A+ d{\bf{x}}_{\perp}^2~,
\label{BRM3}
\end{equation}
where ``$A$'' stands for the transverse indices and $\Theta$ is the Lorentz-Heaviside theta function.

The massless scalar field mode, propagating on the metric (\ref{2.01}), are found by evaluating the solutions of Klein-Gordon equation. They are found to be as follows \cite{Klimcik:1988az,Klimcik:1989kh,Compere:2019rof}:
\begin{equation}
g_{k_-, {\bf k}_{\perp}} = N_ke^{-ik_- V}e^{-ik_+(U-U_0)+i{\bf k}_{\perp}\cdot{\bf x}_{\perp}},~\textrm{for}~ U<U_0~;
\label{2.02}
\end{equation}
and
\begin{eqnarray}
&&f_{k_-, {\bf k}_{\perp}} = N_ke^{-ik_-V}\int \frac{d^2{\bf x'}_{\perp}}{(2\pi)^2} e^{i{\bf k}_{\perp}\cdot {\bf x'}_{\perp}+ik_-f({\bf x'}_{\perp})} 
\nonumber
\\
&&\times \int d^2{\bf k'}_{\perp}e^{i{\bf k'}_{\perp}\cdot {\bf x}_{\perp} - i\frac{{\bf k'}_{\perp}^2}{4k_-} (U-U_0) - i{\bf k'}_{\perp}\cdot {\bf x'}_{\perp}},
\nonumber
\\
&&{\textrm{for}}~U>U_0~.
\label{2.03}
\end{eqnarray}
We call them as shock wave Minkowski mode (SWMM).
The components wave vector are $k^a=(k_-,k_+,{\bf k}_{\perp})$ with $k_{\pm} = (1/2)(k_T\pm k_X)$ and ${\bf k^2}_{\perp} = k_Y^2+k_Z^2$. All the components are not independent and they are related as $k_+=({\bf k}_{\perp}^2/4k_-)$. The normalization is given by $N_k = [(2\pi)^{3/2}\sqrt{2k_-}]^{-1}$. Remember that for positive frequency mode $k_-$ is positive. Now since we have $k_+k_-={\bf k}_\perp^2/4 >0$, then for positive frequency mode $k_+$ has to be positive as well.

The mode (\ref{2.03}) for $U>U_0$ can be cast in a more convenient form if one perform the integration over ${\bf k'}_{\perp}$. This has been explicitly done in \cite{Majhi:2020pps}. The final expression comes out to be
\begin{eqnarray}
&&f_{k_-, {\bf k}_{\perp}} = - N_k\frac{4i\pi k_-}{(U-U_0)}e^{-ik_-V}
\nonumber
\\
&&\times\underbrace{\int \frac{d^2{\bf x'}_{\perp}}{(2\pi)^2} e^{i{\bf k}_{\perp}\cdot {\bf x'}_{\perp}+ik_-f({\bf x'}_{\perp}) + \frac{ik_-|{\bf x}_{\perp}-{\bf x'}_{\perp}|^2}{U-U_0}}}_{I_{\bf x'_\perp}}~.
\label{2.07}
\end{eqnarray}
We shall use this form for our future purpose. 

\section{Defining detector's trajectory}
Suppose the shock wave is localised at $U_0  = -1$ and we want to confine the movement of the detector within the region $U>U_0$. Here  null like trajectory will be obtained.
For simplicity we consider that the detector is moving along a trajectory such that
\begin{equation}
Z=0;\,\,\,\ \hat{X} =0~,
\label{3.01}
\end{equation}
with $U>U_0$.
In this case metric (\ref{BRM3}) reduces to the following form:
\begin{equation}
ds^2  = - d\hat{T}^2 - \frac{d f(Y)}{d Y}d\hat{T}dY + dY^2~.
\label{B1}
\end{equation}
Now since we are interested to null path, set $ds^2=0$ in the above to find the relation between $Y$ and $\hat{T}$ which is followed by the detector. This yields two solutions:
\begin{equation}
\frac{dY}{d\hat{T}} = \frac{1}{2}\Big[\frac{df(Y)}{dY}\pm\sqrt{\Big(\frac{df(Y)}{dY}\Big)^2+4}~\Big]~,
\label{B2}
\end{equation}
where positive indicates outgoing path while negative sign corresponds to ingoing trajectory. Remember that we are going to consider the only ingoing path in our analysis.
This path is chosen in such a way that in both hat and non-hat coordinates the detector is moving along a null path where $U>U_0$. Therefore in this case the trajectory is null like only for $U>U_0$. For the last reason, we had to confine the region of detector's path within $U>U_0=-1$, since the shock is localised at $U=U_0=-1$. In non-hat coordinates the {\it ingoing trajectory} of the detector, by (\ref{BRM2}), is given by
\begin{equation}
T = \hat{T}+\frac{f(Y)}{2}; \,\,\,\,\ X=\frac{f(Y)}{2}; \,\,\,\ Z=0~,
\label{B3}
\end{equation}
where $\hat{T}$ has to be determined in terms of $Y$ from (\ref{B2}).

The present choice of the trajectory is motivated from the following two facts. (a) This already incorporated the classical effect by the shock wave, determined by the shift along $V$ direction in terms of the transverse coordinates. (b) Since we are interested whether classical shock wave can have any effect at the quantum observable, therefore such path may be useful to analyse for quantum signature as it already has the memory of classical shock. With these motivations, and as we already mentioned that the aforesaid trajectory (\ref{B3}) is confined in $U>U_0$ region, here the detector interact only with the modes given by (\ref{2.03}) (or equivalently (\ref{2.07})). Therefore, we shall now aim to investigate if the detector can detect any particles in SWMV when it follows our chosen path.   

In order to proceed further we here choose a particular shock, representing an infinite planar shell of null matter with constant energy density. This is given by \cite{Klimcik:1988az} 
\begin{equation}
f({\bf{x_\perp}}) = -a(Y^2+Z^2)~,
\label{BRM1}
\end{equation}
with $a>0$, in order to energy density is positive. One can choose different shock; but this one is considered here due to its simple structure which makes calculations a bit trivial.
Remember that $f(Y)$ is given by (\ref{BRM1}) with $Z=0$. In this case
the explicit form of the path for our allowed region (i.e. for $U>U_0=-1$), from (\ref{B2}) and (\ref{B3}), turns out to be
\begin{eqnarray}
T &=& -\frac{Y}{2}\sqrt{1+a^2Y^2}-\frac{1}{2a}\sinh^{-1}(aY)~; 
\nonumber
\\
X&=& -\frac{aY^2}{2};
\,\,\,\ Z=0~.
\label{3.02}
\end{eqnarray}
In $(U,V,Y,Z)$ coordinates this is given by
\begin{eqnarray}
&&U = \frac{aY^2}{2} -\frac{Y}{2}\sqrt{1+a^2Y^2}-\frac{1}{2a}\sinh^{-1}(aY)~;
\nonumber
\\
&&V = - \frac{aY^2}{2} -\frac{Y}{2}\sqrt{1+a^2Y^2}-\frac{1}{2a}\sinh^{-1}(aY)~;
\nonumber
\\
&&Z=0~.
\label{3.03}
\end{eqnarray}
Now in order to satisfy the condition $U>U_0=-1$, we choose the range of $Y$ from $-\infty$ to $0$ for the detector's movement. Below in Fig. \ref{Fig1} we show that in this range of $Y$, we have $U>-1$. 
\begin{figure}[!ht]
	\centering
	\includegraphics[scale=0.45, angle=0]{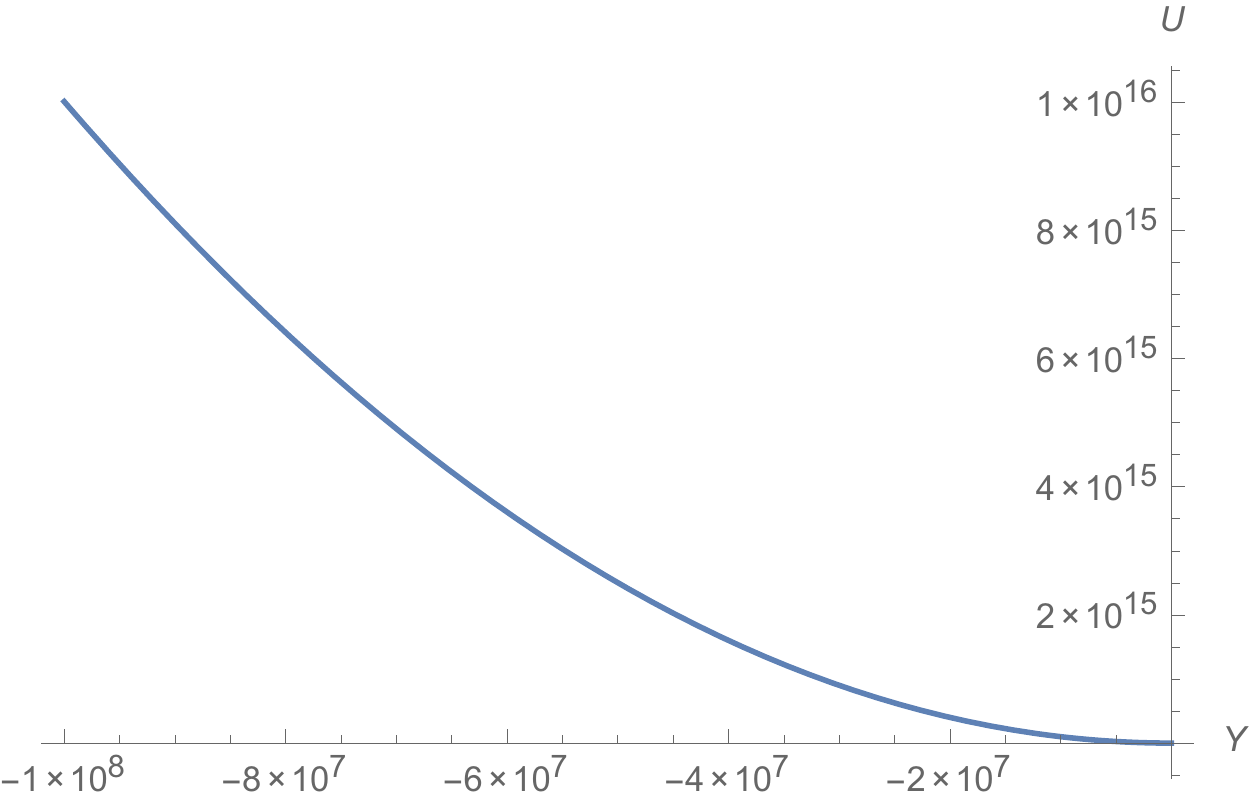}
	\caption{(Color online) Plot for $U$ as a function of $Y$ for $a=1.001$. Here $U$ is positive between $Y=0$ to $Y=-\infty$ and hence $U>U_0=-1$ is satisfied in this range of $Y$.}
	\label{Fig1}
\end{figure}
Therefore this range of detector's movement is well suited for our purpose.

\section{Detector's response}
The detector set up will be identical to one introduced earlier in \cite{Scully:2017utk,Chakraborty:2019ltu} (also see \cite{Dalui:2020qpt}). In brief, the detector is considered to a two level atom. It is interacting with the massless scalar field mode (\ref{2.03}) (or equivalently (\ref{2.07})), since it is confined to move in region denoted by $U>U_0$. In this case the interaction Hamiltonian is considered as
\begin{equation}
H_I = g\Big[(\hat{a}_{k_-,\vec{k}_\perp}f_{k_-, {\bf k}_{\perp}}+h.c.)(\hat{\sigma}_\omega\psi_\omega+h.c.)\Big]~,
\label{4.01}
\end{equation} 
where $\hat{a}_{k_-,\vec{k}_\perp}$ is the scalar field annihilation operator and $\hat{\sigma}_\omega$ is the detector's lowering operator. $\psi_\omega$ is the detector's wave function and $g$ denotes the coupling constant. In this case the transition probability from ground state to excited state of detector, calculated at the first order perturbation series, is given by \cite{Scully:2017utk,Chakraborty:2019ltu}
\begin{equation}
P = g^2 \Big|\int_{\lambda_i}^{\lambda_f} d\lambda f_{k_-, {\bf k}_{\perp}}(\lambda) \psi_\omega(\lambda)\Big|^2~.
\label{4.02}
\end{equation}
In the above $\lambda$ denotes the time of detector's clock.

For the shock wave of type (\ref{BRM1}), the scalar mode (\ref{2.07}) take the following form
\begin{eqnarray}
&&f_{k_-, {\bf k}_{\perp}} = -\frac{e^{-ik_-V}}{{\sqrt{(2\pi)^3 2k_-}\Big[a(U-U_0)-1\Big]}}
\nonumber
\\
&&\times\exp\Big[{\frac{i{\bf k}_\perp^2-\frac{4ik_-}{U-U_0}{\bf k}_\perp\cdot{\bf x}_\perp+\frac{4iak_-^2}{U-U_0}{\bf x}_\perp^2}{4k_-(a-\frac{1}{U-U_0})}}\Big]~.
\label{4.03}
\end{eqnarray}
This can be shown as follows. Substitution of (\ref{BRM1}) in the (\ref{2.07}), the integration $I_{\bf{x'_\perp}}$ yields
\begin{eqnarray}
I_{\bf{x'_\perp}} &=& \frac{1}{(2\pi)^2}\int_{-\infty}^{+\infty} dY' e^{ik_YY'-iak_-Y'^2 + \frac{ik_-}{U-U_0}(Y-Y')^2}
\nonumber
\\
&\times&\int_{-\infty}^{+\infty} dZ' e^{ik_ZZ'-iak_-Z'^2 + \frac{ik_-}{U-U_0}(Z-Z')^2}~.
\label{4.04}
\end{eqnarray}
Here both the integrals are identical. They are evaluated by the following general formula
\begin{equation}
\int_{-\infty}^{+\infty}d\sigma e^{A\sigma-B\sigma^2+C(D-\sigma)^2} = \sqrt{\frac{\pi}{B-C}}~e^{\frac{A^2-4ACD+4BCD^2}{4(B-C)}}~.
\label{4.05}
\end{equation}
This leads to
\begin{eqnarray}
I_{\bf x'_\perp} &=& \frac{1}{4i\pi k_- (a-\frac{1}{U-U_0})} 
\nonumber
\\
&\times&\exp\Big[ \frac{i{\bf k}_\perp^2 - \frac{4ik_-}{U-U_0}{\bf k}_\perp\cdot{\bf x}_\perp + \frac{4iak_-^{2}}{U-U_0}{\bf x}_\perp^2}{4k_-(a-\frac{1}{U-U_0})}\Big]~.
\label{4.06}
\end{eqnarray}  
Using this in (\ref{2.07}) yields our required expression (\ref{4.03}).
Whereas the detector wave function for positive frequency is given by 
\begin{equation}
\psi_\omega = e^{-i\omega \lambda}~.
\label{4.07}
\end{equation}
Using all these we will now calculate the transition probability (\ref{4.02}) for our chosen trajectory.

Here we consider the path of the detector is given by (\ref{3.03}) and its clock is synchronised with time $T$. Therefore we have $\lambda=T$ and now expressing everything in terms of $Y$ from (\ref{3.03}) one finds the transition probability (\ref{4.02}) as
\begin{eqnarray}
&&P= \frac{g^2}{(2\pi)^32k_-}\Big|\int_{-\infty}^{0} dY \Big(\frac{dT}{dY}\Big) \frac{e^{-ik_-V}}{\Big[a(U+1)-1\Big]}
\nonumber
\\
&&\times\exp\Big[{\frac{i({ k}_Y^2+k_Z^2)-\frac{4ik_-}{U+1}{ k}_Y{Y}+\frac{4iak_-^2}{U+1}{Y}^2}{4k_-(a-\frac{1}{U+1})}}\Big]e^{-i\omega (T-i\epsilon)}\Big|^2~,
\nonumber
\\
\label{4.09}
\end{eqnarray}  
where we have used $d\lambda=dT= (dT/dY) dY$ as $T$ is function of $Y$ alone, given in (\ref{3.02}). To compute the above integration one needs to first express $T, U$ and $V$ in terms of $Y$ using (\ref{3.02}) and (\ref{3.03}). Following the earlier argument the integration limits are set to be $-\infty$ to $0$. Also we introduced a very small parameter $\epsilon>0$ to make the integrant convergent, which will be taken to be zero after completion of the integration. 

Since evaluation of the integration in (\ref{4.09}) is analytically not possible  we here compute it numerically. Below in Fig. \ref{Fig2} we numerically plot the rescaled transition probability $R=((2\pi)^32k_-\omega^2 P)/g^2$ as a function of detector's frequency $\omega$ for different values of $a$.
\begin{figure}[!ht]
	\centering
	\includegraphics[scale=0.35, angle=0]{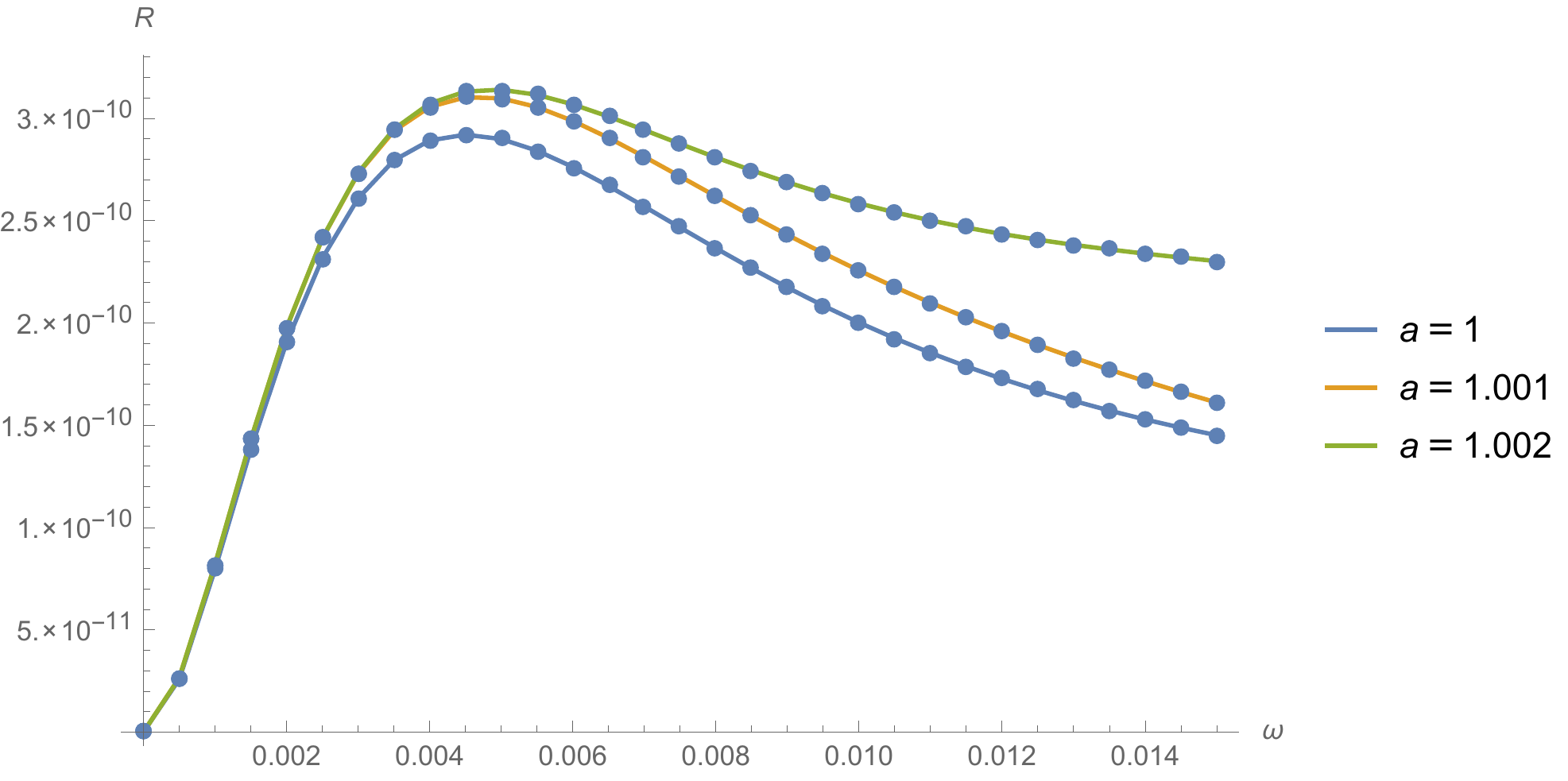}
	\caption{(Color online) $R$ VS $\omega$ plot for different values of $a$. Here we choose $k_- = 9.9$, $k_Y = 10$, $k_Z=17.32$ and $\epsilon = 0.01$.}
	\label{Fig2}
\end{figure}
This shows that for non-vanishing value of $a$ (i.e. in presence of shock wave) the detector measures particle in the SWMV. Moreover, the probability increases with the increase of strength of shock wave, determined by parameter $a$. This last feature (i.e. change of probability with the change in strength of the shock) is important in recognising the presence of shock wave in spacetime through quantum effect.

We finish this section with the following comments.
\begin{itemize}
\item It may be noted that SWMV is defined in the region $U>U_0$ and the scalar field modes incorporate shock information. On the contrary the vacuum and the modes remain unchanged in $U<U_0$ and they are Minkowski vacuum and usual Minkowski modes, respectively. Still the earlier investigations showed that the usual accelerating frame does not see any signature of spacetime shock in SWMV. In the present analysis we particularly interested whether at the quantum level SWMV stores information about the classical shock. Therefore we wanted to concentrate on the mode, given by (\ref{2.03}) and since the mode is only defined on $U>U_0$, we restricts the trajectory of our detector on that side.  Although the trajectory incorporates classical memory of shock and therefore it keeps this forever, even on the other side (i.e. $U<U_0$), but in the latter region the field modes and the corresponding vacuum are different. Hence just to investigate the properties of fields on our interested region, we need to consider the trajectory which is confined there.
\item We found that the detector when moves along our aforesaid trajectory it clicks and the feature of the spectrum depends on the strength of shock. By this latter property we concluded that the quantum memory are now visible. Now if there is no shock in the spacetime (i.e. $a=0$ in Eq. (\ref{BRM1})), then the trajectories (\ref{3.03}) reduces to the following form
\begin{equation}
U=-Y; \,\,\ V=-Y~,
\label{R1}
\end{equation}
with $T=-Y$ and $X=0=Z$. Also the metric (\ref{BRM3}) reduces to the usual Minkowski spacetime. In this case the mode function will be given by of the form (\ref{2.02}) but now it is valid throughout the spacetime. Also the vacuum corresponding is now global Minkowski one. In this case the transition probability (\ref{4.02}) takes the following form
\begin{eqnarray}
P &=& \frac{g^2}{(2\pi)^3 2k_-} \Big|\int_{-\infty}^{\infty} dY \exp[i(\omega + k_T + k_Y)Y]\Big|^2
\nonumber
\\
&=& \frac{g^2}{(2\pi)^3 2k_-} \Big|2\pi \delta(\omega + k_T + k_Y)]\Big|^2~.
\end{eqnarray}
Here we have $k_+,k_->0$ and therefore $k_T=k_++k_->0$. In our earlier numerical analysis we took $k_Y>0$ and hence for $\omega>0$ (as only positive frequency wave function for detector has been chosen here), the above vanishes. Therefore with respect to this detector which measures the transition in absence of shock, our confined detector will conclude that their transition is due to shock only.
\item The trajectory is chosen in such a way that it contains the information about the shock. The null trajectory has been constructed in such a way that in absence of shock, they reduces to constant velocity trajectory (see Eq. (\ref{R1})) which is the known null path in Minkowski spacetime. Hence this trajectory indeed deformed by the shock. The deformation (or shift) can be found by subtracting (\ref{R1}) from (\ref{3.03}).  Our aim was to find a trajectory in natural way on a shock deformed spacetime which are not influenced by any other external agency. In that case the quantum information whatever we calculated must show only the properties of spacetime memory only.

\item Finally it must be mentioned that in order to see the presence of shock at the quantum level we used the well known concept -- the transition in a detector depends on the specific choice of trajectory. In the light of known result that a uniformly accelerated detector does not feel the shock (as the transition probability does not get modified by shock), it is necessary to know how the presence of shock wave can be identified at the quantum level. In that case we here specifically argued that one needs to choose a path which is distorted by shock. Moreover, our chosen trajectory is such that it is trivial in absence of wave. Therefore the detector which is moving along this path shows only the feature of spacetime distortion memory at the quantum level. Any change in quantum results (like transition probability here) due to change in shockwave signifies the quantum memory effect which was not captured in earlier discussions. Hence we conclude that our present detector's response due to choice of specific paths is not a novel idea; rather we have used this fact to identify trajectories which are well capable of revealing the spacetime memory effect at the quantum level.  
\end{itemize}


\section{Conclusion} 
The ongoing investigations on the quantum memory, seen by observers, due to shock wave on spacetime is being continued here. It was observed in \cite{Compere:2019rof} that a usual accelerated observer does not see any modification to Unruh spectrum for the Dray and 't Hooft spacetime \cite{Dray:1984ha}. Moreover, since the Wightman function is unaltered \cite{Majhi:2020pps}, as a result any known observer, like rotating frame, accelerated frame etc, dose not see any signature of it. Here, within Dray - 't Hooft setup,  we argue that this is because the observers' trajectories do not have any classical memory of shock wave. In fact we need to consider those class of frames which are also affected by it along with the underlying spacetime. The underlying reason is as follows. Due to the shock, the transformations from the affected Minkowski frame to our required frame should also incorporate the information, as like the scalar modes are being modified. This has not been considered in the earlier analysis.

In this short note we found a null like observer whose path is affected by the shock wave. Here it has been observed that a two level atomic detector, moves along this trajectory, clicks and can see particle in the SWMV. We showed that the probability of transition within the detector's quantum levels depends on the strength of shock in the spacetime. Therefore the measure of transition probability, when a detector moves along this particular path, can be a very good candidate to indicate the shock memory at the quantum level. Hence we conclude that the quantum memory of shock wave can be very specific to certain class of observers. Hope this piece of information will be important for further progress of present area of investigation.


\vskip 5mm
{\it Acknowledgments}: 
I thank anonymous referees for their useful comments. The research of the author is supported by a START- UP RESEARCH GRANT (No. SG/PHY/P/BRM/01) from the Indian Institute of Technology Guwahati, India and by a Core Research Grant (File no. CRG/2020/000616) from Science and Engineering Research Board (SERB), Department of Science $\&$ Technology (DST), Government of India.
 
{\sc This work is dedicated to those who are helping us to fight against COVID-19 across the globe.}

\
\end{document}